\documentclass[oldversion]{aa}  
\usepackage{graphicx}
\usepackage{txfonts}
\begin{document}

\title{Open cluster stability and the effects of binary stars}
\titlerunning{Open cluster stability and binary stars}

\author{R. de Grijs,$^{1,2}$ S. P. Goodwin,$^1$ M. B. N. 
Kouwenhoven,$^1$ \and P. Kroupa$^3$}
\authorrunning{R. de Grijs et al.}
\offprints{R. de Grijs}

\institute{$^1$ Department of Physics \& Astronomy, The University of
Sheffield, Hicks Building, Hounsfield Road, Sheffield S3 7RH, UK\\
\email{[R.deGrijs; S.Goodwin; T.Kouwenhoven]@sheffield.ac.uk}\\
$^2$ National Astronomical Observatories, Chinese Academy of Sciences,
20A Datun Road, Chaoyang District, Beijing 100012, China\\
$^3$ Argelander-Institut f\"ur Astronomie, Universit\"at Bonn, Auf dem
H\"ugel 71, D-53347 Bonn, Germany\\
\email{pavel@astro.uni-bonn.de}}

\date{Received / accepted}

\abstract{The diagnostic age versus mass-to-light ratio diagram is
often used in attempts to constrain the shape of the stellar initial
mass function (IMF), and the potential longevity of extragalactic
young to intermediate-age massive star clusters. Here, we explore its
potential for Galactic open clusters. On the basis of a small,
homogenised cluster sample we provide useful constraints on the
presence of significant binary fractions. Using the massive young
Galactic cluster Westerlund 1 as a key example, we caution that
stochasticity in the IMF introduces significant additional
uncertainties. We conclude that for an open cluster to survive for any
significant length of time, and in the absence of substantial external
perturbations, it is a necessary but not a sufficient condition to be
located close to or (in the presence of a significant binary
population) somewhat {\it below} the predicted photometric
evolutionary sequences for `normal' simple stellar populations
(although such a location may be dominated by a remaining `bound'
cluster core and thus not adequately reflect the overall cluster
dynamics).}


\maketitle

\section{Introduction}
\label{intro.sec}

Over the past few years, detailed studies of the stellar content and
longevity of {\it extragalactic} massive star clusters have
increasingly resorted to the use of the age versus mass-to-light
($M/L$) ratio diagram as a diagnostic tool, where one usually compares
dynamically determined $M/L$ ratios with those predicted by the
evolution of `simple' stellar populations\footnote{As a `normal' SSP
we define a coeval stellar population of a single metallicity and
characterised by either a Salpeter (1955) or a Kroupa (2001)-type
stellar initial mass function (IMF), i.e., a two-part power law
covering the stellar mass range from $0.1 M_\odot$ to $\sim 125
M_\odot$, depending on metallicity.} (SSPs; e.g., Smith \& Gallagher
2001; Mengel et al. 2002; McCrady, Gilbert \& Graham 2003; Larsen et
al. 2004; McCrady, Graham \& Vacca 2005; Bastian et al. 2006; Goodwin
\& Bastian 2006; de Grijs \& Parmentier 2007; Moll et al. 2008). Based
on high-resolution spectroscopy to obtain the objects' line-of-sight
(1D) velocity dispersions, $\sigma$, and on high spatial resolution
imaging to obtain accurate, projected half-light
radii\footnote{Assuming that light traces mass, the observed
half-light radii must be corrected for projection onto the sky by
applying a correction factor of 3/4 (e.g., Fleck et al. 2006).},
$r_{\rm hl}$, most authors then calculate the dynamical cluster
masses, $M_{\rm dyn}$, using
\begin{equation}
\label{spitzer.eq}
M_{\rm dyn} = \eta \frac{r_{\rm hl} \sigma^2}{G} \quad ,
\end{equation}
where $G$ is the gravitational constant (Limber \& Mathews 1960; see
also Aarseth \& Saslaw 1972; Spitzer 1987). In Eq. (\ref{spitzer.eq}),
$\eta \approx 9.75$ is a dimensionless parameter which is usually
assumed to be constant (but see Fleck et al. 2006; Kouwenhoven \& de
Grijs 2008). The dominant assumptions underlying the validity of
Eq. (\ref{spitzer.eq}) are that the cluster is in virial equilibrium,
and that it consists of {\em single} stars of equal mass. While the
latter assumption introduces an offset in the cluster mass of only
approximately a factor of two compared to using a reasonable range of
stellar masses (e.g., Mandushev, Spassova \& Staneva 1991; see also
Fleck et al. 2006, and references therein), the former breaks down for
ages younger than about 15 Myr. In reality, the effects of mass
segregation (Fleck et al. 2006) and a high fraction of binary and
multiple systems (Kouwenhoven \& de Grijs 2008) also significantly
affect the total cluster mass estimates obtained from integrated
(whole-cluster) velocity dispersion measurements.

Nevertheless, using this approach one can get at least an initial
assessment as to whether a given (unresolved) cluster may be (i)
significantly out of virial equilibrium, in particular `super-virial',
(ii) substantially over- or underabundant in low-mass stars, or (iii)
populated by a large fraction of binary and higher-order multiple
systems. Since the work by Bastian \& Goodwin (2006) and Goodwin \&
Bastian (2006), we can now also model any (super-virial) deviations
from the SSP models for the youngest ages (up to $\sim 40$ Myr) if we
assume that these are predominantly due to clusters being out of
virial equilibrium after gas expulsion.

This has led a number of authors to suggest that, in the absence of
significant external perturbations, massive clusters located in the
vicinity of the SSP models and aged $\ga 10^8$ yr may survive for a
Hubble time and eventually become old globular cluster (GC)-like
objects (e.g., Larsen et al. 2004; Bastian et al. 2006; de Grijs \&
Parmentier 2007).

Encouraged by the recent progress in this area based on both
observational and theoretical advances, in this paper we set out to
address the following unresolved questions:
\begin{enumerate} 
\item {\it Can we use the integrated velocity dispersions of
extragalactic massive star clusters as valid proxies of their
gravitational potential?} To address this question we need to have
access to the kinematics (as well as other physical properties) of the
individual stars in a given set of sufficiently massive clusters. At
present, the only massive cluster for which such data are available is
the Galactic cluster Westerlund 1 (Section \ref{wld1.sec}). Resolving
this issue robustly is of prime importance in the field of
extragalactic massive star cluster research, where it is implicitly
assumed yet untested.

\item {\it Can we use the internal kinematics of Galactic open
clusters to constrain their binary stellar populations in a
straightforward manner?} To gain physical insights into this open
issue crucial for stellar population modelling we obtained the
relevant observations for a small sample of nearby open clusters for
which proper motions and radial velocities were readily available for
the individual member stars (Section \ref{data.sec}). We derive the
(apparent) dynamical $M/L$ ratios and discuss their implications in
Sections \ref{data.sec} and \ref{disc.sec}. We particularly focus on
the application of the new binary diagnostic diagram proposed by
Kouwenhoven \& de Grijs (2008).

\item {\it Is it sufficient for a cluster, based on resolved kinematic
measurements, to be located close to the SSP model predictions in
order for it to survive for any significant length of time, in the
absence of external perturbations?} Given that this is a key
underlying assumption in many extragalactic massive cluster studies,
our approach in Sections \ref{disc.sec} and \ref{summary.sec} is to
revert this question: are there clear examples of clusters that
satisfy this condition yet are known to be in the final stages of
dissolution? If we can answer `yes' to the latter question, this will
serve as a clear caution, as emphasised in Section \ref{summary.sec},
where we summarise our main results, cautions and conclusions.
\end{enumerate}

\section{Observational data}
\label{data.sec}

\begin{figure}
\includegraphics[width=\columnwidth]{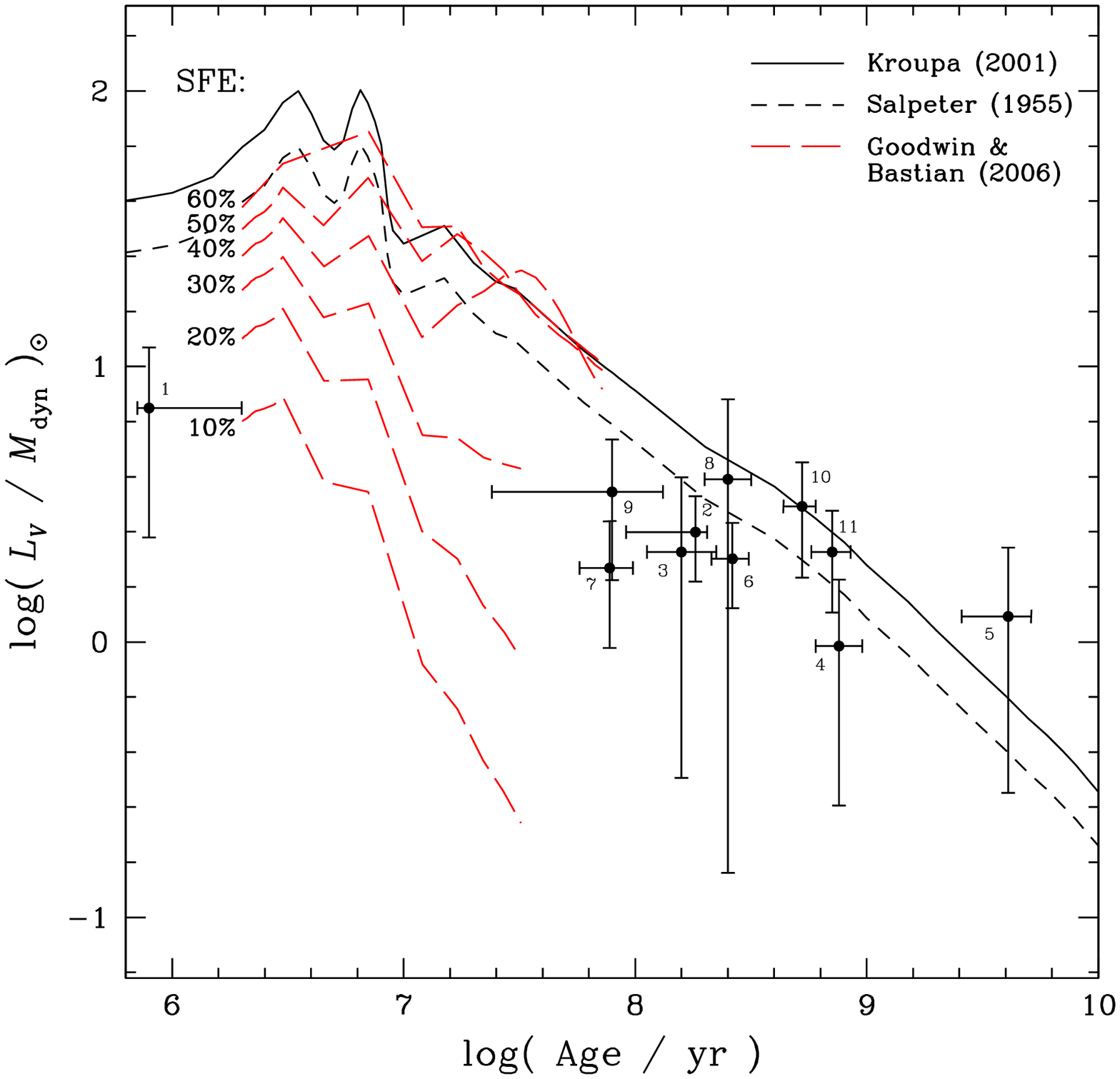}
\caption{\label{ocl.fig}Diagnostic age versus $M/L_V$ ratio diagram,
including the Galactic open clusters for which velocity dispersion
measurements are available. The evolution expected for SSPs governed
by both Salpeter (1955) and Kroupa (2001) IMFs is shown as the solid
and short-dashed lines, respectively. The long-dashed lines represent
the evolution expected for SSPs with a Kroupa (2001)-type IMF, but a
range of effective star-formation efficiencies (Goodwin \& Bastian
2006). The sizes of the error bars are based on the most realistic
ranges of observable values used to calculate the clusters' loci in
this diagram.\hfill\eject Numbered clusters: 1, NGC 1976 (Orion Nebula
Cluster); 2, NGC 2168; 3, NGC 2516; 4, NGC 2632; 5, NGC 2682; 6, NGC
3532; 7, NGC 5662; 8, NGC 6705; 9, Pleiades; 10, Coma Berenices; 11,
Hyades.}
\end{figure}

In order to test the usefulness of the diagnostic diagram of cluster
age versus $M/L$ ratio for Galactic open clusters (see
Fig. \ref{ocl.fig}, which we will discuss in detail in Section
\ref{disc.sec}), we rely on published parameters. Since each of the
observables has an associated uncertainty, it is paramount that we
base our results on data sets that are as homogeneous as possible. The
most crucial ingredient for the dynamical $M/L$ ratio determination is
the internal velocity dispersion. We include only those Galactic open
clusters for which these velocity dispersions have been derived from
the proper motions of the individual stars\footnote{Although, strictly
speaking, Eq. (1) is valid for line-of-sight velocity dispersions
instead of the equivalent dispersions based on proper-motion studies,
the effect of ignoring this will be mass overestimates by a factor of
2 (or $\Delta \log L/M = +0.3$) if the clusters' kinematics are
isotropic. This does not affect our conclusions.} (for NGC 3532 the
internal velocity dispersion used in this paper is based on individual
stellar radial velocity measurements; Gieseking 1981). Where possible
we include the {\it core} velocity dispersions, in order to match the
structural parameters we will use. We also require well-determined
distances (to obtain luminosities and linear velocity dispersions) as
well as core radii and photometric observables. The distance estimates
used here are mostly based on the recent homogenised compilations of
Kharchenko et al. (2005) and Dias et al. (2006), supplemented with
determinations based on a number of studies focusing on individual
clusters. Although Kharchenko et al. (2005) provide values for the
core radii of many of our sample clusters, the associated
uncertainties are large. In fact, they often dominate our dynamical
mass estimates, together with the often large uncertainties in the
integrated $V$-band magnitudes. The latter are often difficult to
obtain to any reasonable degree of accuracy because of the crowded
fields in which many of the clusters are located and also because of
uncertain stellar cluster membership determinations (see also the
discussion in Section \ref{wld1.sec}).

Nevertheless, in Table \ref{observables.tab} we have collected the
`best' values for the core radii, $R_{\rm c}$, distances, $D$,
apparent $V$-band magnitudes, foreground extinction, $E(B-V)$, and
velocity dispersions, $\sigma$. We also list their likely
(uncertainty) ranges for our sample clusters. We provide for both the
values and the uncertainties the references we have used, and we have
aimed to homogenise our cluster sample parameters (following a similar
procedure as Paunzen \& Netopil 2006, although they used different
selection criteria). This implies that our choice of the `best' values
for certain parameters may depend on the values of one or more of the
other observables. We provide the full list of references used to
obtain the most likely parameter ranges. However, where we have
discarded certain values (often because they were clear statistical
outliers), the relevant respective references are bracketed.

In Table \ref{derived.tab} we list the best ages and their uncertainty
ranges of our sample clusters using the same notation as in Table
\ref{observables.tab}, as well as the total cluster masses -- based on
Eq.~(\ref{spitzer.eq}), with $\eta = 9.75$ -- and their $L_V/M$ ratios
derived based on the parameters and their ($\sim$1$\sigma$)
uncertainties from Table \ref{observables.tab}. A full list of
references to Tables \ref{observables.tab} and \ref{derived.tab} is
provided in Table \ref{refs.tab}.

We note that our sample selection is biased towards the nearest
Galactic open clusters, for which reasonably accurate internal
velocity dispersions could be obtained. However, although our sample
is by no means complete in any sense, we can still use it to assess
(i) the binary fractions of the clusters individually (Section
\ref{disc.sec}) and (ii) the usefulness of the diagnostic age versus
$M/L$ ratio diagram for cluster longevity considerations (Section
\ref{summary.sec}).

\begin{table*}
\caption[ ]{\label{observables.tab}Observational parameters of the open cluster sample}
\begin{center}
{\scriptsize
\tabcolsep=1.5pt
\begin{tabular}{lclclrlrllllllllllllll}
\hline
\hline
Name     & \multicolumn{2}{c}{$R_{\rm c}$} & \multicolumn{2}{c}{$\pm$} & \multicolumn{2}{c}{$D$} & \multicolumn{2}{c}{$\pm$} & \multicolumn{2}{c}{$V$} & \multicolumn{2}{c}{$\pm$} & \multicolumn{2}{c}{$E(B-V)$} & \multicolumn{2}{c}{$\pm$} & \multicolumn{2}{c}{$\sigma$} & \multicolumn{2}{c}{$\pm$} \\
         & \multicolumn{1}{c}{(arcmin)} & \multicolumn{1}{c}{ref.} & \multicolumn{1}{c}{(arcmin)} & \multicolumn{1}{c}{ref.} & \multicolumn{1}{c}{(pc)} & \multicolumn{1}{c}{ref.} & \multicolumn{1}{c}{(pc)} & \multicolumn{1}{c}{ref.} & \multicolumn{1}{c}{(mag)} & \multicolumn{1}{c}{ref.} & \multicolumn{1}{c}{(mag)} & \multicolumn{1}{c}{ref.} & \multicolumn{1}{c}{(mag)} & \multicolumn{1}{c}{ref.} & \multicolumn{1}{c}{(mag)} & \multicolumn{1}{c}{ref.} & \multicolumn{1}{c}{(km s$^{-1}$)} & \multicolumn{1}{c}{ref.} & \multicolumn{1}{c}{(km s$^{-1}$)} & \multicolumn{1}{c}{ref.} \\
\hline
NGC 1976 &  7.2  & 38 & 2.4  & 13    &  470 & 13 &  70 & 9,20,45     & 1.26 & 21    & 0.1 & $^a$  & 0.05 & 9,20  & 0.01 & $^a$          & 2.34 & 45 & 0.7  & 17,45 \\
         &       &    &      &       &      & 38 &     &             &      &       &     &       &      &       &      &               &      &    &      &    \\
NGC 2168 & 12.0  & 20 & 1.2  & $^a$  &  912 &  9 &  68 & 3,9,19,20   & 4.86 & 42,43 & 0.2 & $^a$  & 0.20 & 9,19  & 0.05 &  3,20,35,39   & 1.00 & 26 & 0.10 & 26 \\
         &       &    &      &       &      & 19 &     & 26,42,43    &      &       &     & (3)   &      &       &      & 42,43,47      &      &    &      &    \\
NGC 2516 & 12.0  & 20 & 1.8  & $^a$  &  358 & 41 &  60 &  3,6,9      & 3.1  & $^b$  & 0.3 &  3,35 & 0.11 & 41    & 0.02 & 7,9,20,35     & 1.35 & 33 & 0.35 & 33 \\
         &       &    &      &       &      &    &     & 20,42,43    &      &       &     &       &      &       &      & 41,42,43      &      &    &      &    \\
NGC 2632 & 66.0  & 15 & 4.0  & 20    &  171 & 31 &  12 &  2,9,20     & 3.2  & 39    & 0.1 &  3,35 & 0.00 & 31,35 & 0.01 & 9,20,36       & 0.48 & 16 & 0.2  & 16 \\
         &       &    &      &       &      &    &     & 31          &      &       &     &       &      & 39    &      &               &      &    &      &    \\
NGC 2682 &  4.7  &  4 & 0.6  &  4    &  820 & 31 &  47 & 31,36,37    & 6.5  & 36    & 0.3 & 35,39 & 0.05 & 31    & 0.02 & 9,20,31,35,36 & 0.81 & 12 & 0.4  & 12,25 \\
         &       &    &      &       &      &    &     & 42,43       &      &       &     & 42,43 &      &       &      & 39,42,43,44   &      &    &      &    \\
NGC 3532 & 12.0  & 20 & 1.2  & $^a$  &  492 & 31 &   8 & 3,9,20,31   & 3.1  &  3    & 0.2 & $^a$  & 0.04 &  3,20 & 0.01 & 5,9,31        & 1.49 & 11 & 0.29 & 11 \\
         &       &    &      &       &      &    &     &             &      &       &     &       &      & 31    &      &               &      &    &      &    \\
NGC 5662 &  6.0  & 20 & 1.2  & $^a$  &  684 & 31 &  60 & 9,20,31,34  & 5.6  & 30    & 0.2 & $^a$  & 0.32 & 30,31 & 0.01 & 9,20,31,34    & 1.2  & 33 & 0.3  & 33 \\
         &       &    &      &       &      &    &     & 42,43       &      &       &     &       &      & 42,43 &      &               &      &    &      &    \\
NGC 6705 &  1.38 & 27 & 0.90 &  8,20 & 2042 & 40 & 150 &  3,8,9,20   & 6.8  & 27    & 0.4 &  3,42 & 0.43 & 9,20  & 0.03 & 3,8,30,39,40  & 2.0  & 26 & 0.8  & 24,25 \\
         &       &    &      & 23,27 &      &    &     & 27,40,42,43 &      &       &     & 43    &      & 40    &      & 42,43         &      &    &      &    \\
Pleiades & 66.0  &  1 & 9.0  &  1    &  133 & 31 &   9 & 9,20,31     & 1.2  & 35    & 0.1 & 39    & 0.05 & 31    & 0.01 & 9,20,31,35,39 & 0.6  & 26 & 0.2  & 15 \\
         &       & 20 &      &       &      &    &     &             &      &       &     &       &      &       &      &               &      &    &      &    \\
Coma Ber & 96.0  & 28 & 9.0  & 20,28 &   86 & 31 &   7 & 9,20,28,31  & 1.8  & 35    & 0.2 & $^a$  & 0.00 & 20,31 & 0.01 & 9,31          & 0.3  & 28 & 0.2  & $^a$ \\
         &       &    &      &       &      &    &     &             &      &       &     &       &      & 35    &      &               &      &    &      &    \\
Hyades   &205.8  & 32 & 2.0  & $^a$  &   42 & 31 &   3 & 9,31        & 0.5  & $^b$  & 0.1 & 35,39 & 0.00 & 31,35 & 0.01 & 9,31          & 0.32 & 22 & 0.2  & $^a$ \\
         &       &    &      &       &      &    &     &             &      &       &     &       &      & 36,39 &      &               &      &    &      &     \\
\hline
\end{tabular}
} 
\end{center}
\flushleft 
{\sc Notes:} $^a$ adopted based on realistic measurement errors for
large samples of open clusters (Dias et al. 2002; and unpublished);
$^b$ average
\end{table*}

\begin{table*}
\caption[ ]{\label{derived.tab}Derived parameters of the open cluster sample}
\begin{center}
\begin{tabular}{lclclcc}
\hline
\hline
\multicolumn{1}{c}{Name} & \multicolumn{1}{c}{$\log (\mbox{Age
yr}^{-1})$} & \multicolumn{1}{c}{ref.} & \multicolumn{1}{c}{$\pm$} &
\multicolumn{1}{c}{ref.} & \multicolumn{1}{c}{$M_{\rm dyn} ({\rm
M}_\odot)$} & \multicolumn{1}{c}{$\log( L_V/M_{\rm dyn}) [{\rm
L}_{V,\odot}/{\rm M}_\odot]$} \\
\hline
NGC 1976 & 5.90 & 38    & $^{+0.40}_{-0.05}$ & (9,20),29         & $9400 \pm 5250$ &  $0.85^{+0.22}_{-0.47}$ \\
NGC 2168 & 8.26 & 9,18  & $^{+0.05}_{-0.30}$ & 3,20,39,42,43,47  & $5551 \pm 1047$ &  $0.40^{+0.13}_{-0.18}$ \\
NGC 2516 & 8.2  & 41    & 0.15               & 3,6,9,20,(42,43)  & $3970 \pm 1680$ &  $0.33^{+0.27}_{-0.82}$ \\
NGC 2632 & 8.88 & 31    & 0.10               & 2,9,20,31,36,(39) & $1319 \pm  920$ &  $0.02^{+0.25}_{-0.59}$ \\
NGC 2682 & 9.61 & 31    & $^{+0.10}_{-0.20}$ & 9,10,14,20,31,36  & $1277 \pm  907$ &  $0.10^{+0.25}_{-0.64}$ \\
         &      &       &                    & 37,39,42,43,46    & $             $ &                         \\
NGC 3532 & 8.42 & 31    & $^{+0.07}_{-0.09}$ &  3,5,9,20,31      & $6649 \pm 1950$ &  $0.30^{+0.13}_{-0.18}$ \\
NGC 5662 & 7.89 & 31    & $^{+0.10}_{-0.13}$ &  9,20,30,31       & $2998 \pm 1246$ &  $0.27^{+0.17}_{-0.29}$ \\
         &      &       &                    & 34,42,43          & $             $ &                         \\
NGC 6705 & 8.4  & 9,40  & 0.1                &  3,(8),20,27      & $5717 \pm 4956$ &  $0.59^{+0.29}_{-1.43}$ \\
         &      &       &                    & (30,39),42,43     & $             $ &                         \\
Pleiades & 7.90 & 31    & $^{+0.22}_{-0.52}$ & 9,20,31,39        & $1603 \pm  794$ &  $0.54^{+0.19}_{-0.32}$ \\
Coma Ber & 8.72 & 31    & $^{+0.06}_{-0.08}$ & 9,20,28,31        & $ 377 \pm  112$ &  $0.49^{+0.16}_{-0.26}$ \\
Hyades   & 8.85 & 31,36 & $^{+0.08}_{-0.09}$ & 9,31,39           & $ 450 \pm  170$ &  $0.33^{+0.15}_{-0.22}$ \\
\hline
\end{tabular}
\end{center}
\end{table*}

\begin{table*}
\caption[ ]{\label{refs.tab}References to Tables \ref{observables.tab}
and \ref{derived.tab}}
{\scriptsize
\begin{center}
\tabcolsep=4pt
\begin{tabular}{rlclclcl}
\hline
\hline
No. & Reference & No. & Reference & No. & Reference & No. & Reference \\
\hline
 1 & Adams et al. (2001)         & 13 & Hillenbrand \& Hartmann (1998) & 25 & McNamara \& Sanders (1977)   & 37 & Sandquist (2004)             \\
 2 & Adams et al. (2002)         & 14 & Hurley et al. (2005)           & 26 & McNamara \& Sekiguchi (1986) & 38 & Scally et al. (2005)         \\
 3 & Batinelli et al. (1994)     & 15 & Jones (1970)                   & 27 & Nilakshi et al. (2002)       & 39 & Spassova \& Baev (1985)      \\
 4 & Bonatto \& Bica (2003)      & 16 & Jones (1971)                   & 28 & Odenkirchen et al. (1998)    & 40 & Sung et al. (1999)           \\
 5 & Clari\'a \& Lapasset (1988) & 17 & Jones \& Walker (1988)         & 29 & Palla \& Stahler (1999)      & 41 & Sung et al. (2002)           \\
 6 & Dachs \& Kabus (1989)       & 18 & Joshi \& Sagar (1983)          & 30 & Pandey et al. (1989)         & 42 & Tadross (2001)               \\
 7 & Dambis (1999)               & 19 & Kalirai et al. (2003)          & 31 & Paunzen \& Netopil (2006)    & 43 & Tadross et al. (2002)        \\
 8 & Danilov \& Seleznev (1994)  & 20 & Kharchenko et al. (2005)       & 32 & Perryman et al. (1998)       & 44 & Taylor (2007)                \\
 9 & Dias et al. (2002)          & 21 & Kopylov (1952)                 & 33 & Sagar \& Bhatt (1989)        & 45 & van Altena et al. (1988)     \\
10 & Dinescu et al. (1995)       & 22 & Makarov et al. (2000)          & 34 & Sagar \& Cannon (1997)       & 46 & VandenBerg \& Stetson (2004) \\
11 & Gieseking (1981)            & 23 & Mathieu (1984)                 & 35 & Sagar et al. (1983)          & 47 & von Hippel et al. (2002)     \\
12 & Girard et al. (1989)        & 24 & Mathieu (1985)                 & 36 & Salaris et al. (2004)        \\
\hline
\end{tabular}
\end{center}
}
\end{table*}

\section{The dynamical state of Galactic open clusters}
\label{disc.sec}

\subsection{Open clusters in the diagnostic diagram}
\label{general.sec}

Using the observational data from Section \ref{data.sec}, we applied
Eq. (\ref{spitzer.eq}) to derive the dynamical masses for each of our
sample cluster cores and calculated the relevant $M/L_V$ ratios. Their
loci in the diagnostic diagram are shown in Fig. \ref{ocl.fig}.
Overplotted is the expected evolution of SSPs (Maraston 2005) for both
a Salpeter (1955) and a Kroupa (2001) stellar IMF (solid and
short-dashed lines, respectively).

We have also included the expected evolution of clusters formed with a
variety of {\it effective} star-formation efficiencies (eSFEs; Goodwin
\& Bastian 2006). The eSFE is a measure of the extent to which a
cluster is out of equilibrium after gas expulsion, on the basis that
the virial ratio {\it immediately before} gas expulsion was $Q_{\rm
vir} = T/|\Omega| = 1/2$(eSFE) (where $T$ and $\Omega$ are the kinetic
and potential energy of the stars, respectively, and a system in
virial equilibrium has $Q_{\rm vir} = 1/2$). The eSFE corresponds to
the true SFE if the stars and gas were initially in virial equilibrium
(see Goodwin \& Bastian 2006).

Owing to the nature of our sample, only\footnote{Despite the extent of
the error bar associated with the age estimate of the Pleiades, Kroupa
et al. (2001) showed this cluster to have re-virialised by an age of
50 Myr, so that it is unlikely affected by the aftermath of the
gas-expulsion phase.} the Orion Nebula Cluster (ONC; cluster 1 in
Fig. \ref{ocl.fig}) is currently young enough so as to possibly be
affected by the effects of rapid gas expulsion, as shown by the extent
(in terms of age) of the long-dashed lines in Fig. \ref{ocl.fig}. The
majority of our sample clusters are old enough ($\ga 40$ Myr) to have
re-virialised after gas expulsion. The dynamical state of these
objects is therefore dominated by the combined effects of (internal)
two-body relaxation, binary motions, and external perturbations.

The fact that our sample of surviving open cluster cores lie close to
the SSP predictions should be expected. Clusters significantly below
the SSP lines will be dynamically `hot' and are expected to dissolve
rapidly, whilst clusters significantly above the lines will be
dynamically `cold' and should (re-)virialise over a few crossing times
to move closer to the canonical SSP lines.

We will now explore the reasons as to why most of our sample clusters
(cores) are found somewhat below the SSP model curves (i.e., they seem
somewhat supervirial with respect to the expectations from the SSP
models), irrespective of whether or not they actually follow the SSP
predictions or are characterised by roughly constant $M/L$ ratios as a
function of age. We expect errors in the core radii to be random, and
unbiased by the mass of a cluster. However, the use of the {\it core}
velocity dispersions and radii may introduce a systematic bias in the
dynamical mass estimates.

There are three dynamical effects that could affect the position of
the clusters relative to the SSP predictions.

{\it First, equipartion and mass segregation could lower the core
velocity dispersion relative to the `typical' velocity and hence move
the clusters' positions to above the SSP predictions.} The majority of
the star clusters in our sample are older than $\sim 10^8$ yr, which
implies that they have ages greater than their half-mass relaxation
times (see, e.g., Danilov \& Seleznev 1994 for the relevant
time-scales for most of our sample clusters). Therefore, these
clusters (and particularly their cores) are expected to be close to
energy equipartition, and thus are significantly mass segregated -- as
observed for, e.g., NGC 2168 (Sung \& Bessell 1999; Kalirai et
al. 2003), NGC 2682 (Bonatto \& Bica 2003), and NGC 6705 (e.g., Sung
et al. 1999 and references therein) among our present sample.
Equipartition reduces the {\it global} velocity dispersion of
high-mass stars relative to low-mass stars, causing high-mass stars to
migrate to the cluster core. Therefore, we might expect the core
velocity dispersion of low-mass cores (as characteristic for the open
clusters discussed in this paper) to underestimate the dynamical mass
of the clusters as a whole and thus produce colder
clusters\footnote{For a quantitative estimate of this effect, let us
assume that our clusters are well represented by Plummer
models. However, we note that this is an unproven assumption; younger
clusters are likely more extended (e.g. Elson, Fall \& Freeman 1987),
whereas older clusters (particularly lower-mass objects) may be
significantly depleted in their outer regions and hence could be much
more compact. A back-of-the-envelope calculation shows then that the
following relations apply (from Heggie \& Hut 2003): $R_{\rm c,intr} =
R_{\rm hm,proj} / \sqrt{2}$, $R_{\rm virial} = R_{\rm hm,proj} \times
16 / 3 \pi$, and $R_{\rm hm,intr} \simeq 1.305 R_{\rm hm,proj}$. This
leads, approximately, to $R_{\rm c,intr} \simeq 1.035 R_{\rm c,proj}$,
and therefore $R_{\rm hm,proj} \simeq 1.464 R_{\rm c,proj}$. Here, the
subscripts `c', `hm', `intr', and `proj' stand for core, half mass,
intrinsic, and projected. This result only holds {\it approximately}
for a Plummer model; it gives us a rough idea of the errors involved
in our analysis, leading to $\Delta(L_V/M_{\rm dyn}) \sim -0.165$ (in
solar units).} -- as seen in Fig. \ref{ocl.fig} (although we remind
the reader of the expected re-virialisation discussed above; this may
introduce an observational bias in the sense that we would not be able
to detect low-mass cluster cores that are significantly super-virial
and hence -- possibly -- in the process of dissolution).

{\it Secondly, the clusters' mass functions (MFs) will have been
altered by dynamical evolution, with the preferential loss of low-mass
stars moving clusters to above the SSP predictions.} This will result
in a `top-heavy' MF in clusters, which will in turn lead to lower
$M/L_V$ ratios than would be expected from the canonical SSP
models. The degree to which the MF will change depends on the two-body
relaxation time which, to first order, depends on the mass of the
cluster (and also on its size; however, we ignore this for now). Thus,
old low-mass clusters are expected to have top-heavy MFs compared to
old high-mass clusters. Therefore, we would expect low-mass clusters
to lie some way above the canonical SSP models, and high-mass clusters
to lie slightly above these lines.\footnote{However, we need an
unbiased sample to explore this option statistically and in more
detail.}

\begin{figure}
\includegraphics[width=\columnwidth]{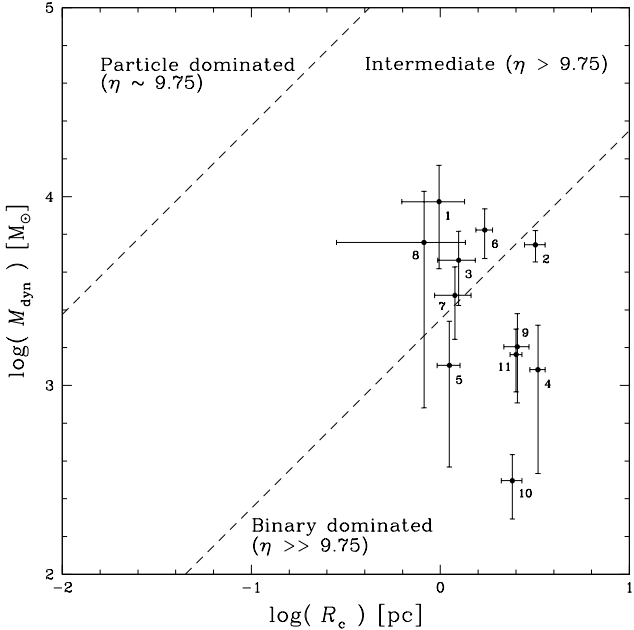}
\caption{\label{fig9.fig}Diagnostic $R_{\rm c}$ versus $M_{\rm dyn}$
diagram, following Kouwenhoven \& de Grijs (2008), including our
sample open clusters. The basic data used to generate the data points
and their uncertainties is included in Tables \ref{observables.tab}
and \ref{derived.tab}; the cluster numbers correspond to those in
Fig. \ref{ocl.fig}.}
\end{figure}

{\it Thirdly, the presence of binaries may result in an observed
velocity dispersion that is higher than the `true' value, moving the
clusters to below the SSP predictions.} Kouwenhoven \& de Grijs (2008)
pointed out that if the velocity dispersion of binary systems was
similar to the velocity dispersion of the cluster (core) as a whole,
the {\it observationally measured} velocity dispersion would
overestimate the mass of a cluster. We can explore, to first order,
whether the binary population may be a significant factor causing an
offset in Fig.~\ref{ocl.fig} by using the new diagnostic proposed by
Kouwenhoven \& de Grijs (2008; their fig. 9). In Fig. \ref{fig9.fig}
we reproduce the main features of their fig. 9, and include our open
cluster sample (using the clusters' core radii instead of their
half-mass radii; the core radii are more likely to represent the size
of the bound stellar population for these clusters; see, e.g.,
Odenkirchen, Soubiran, \& Colin 1998 for arguments relating to the
open cluster in Coma Berenices). It is immediately clear from the
location of the data points that the vast majority of our sample
clusters are indeed expected to be significantly affected by binaries
($\eta > 9.75$, cf. Fig. \ref{fig9.fig}; in fact, the data points
represent upper limits to the cluster masses given that we do not know
the intrinsic masses but need to rely on dynamical tracers). This
seems to be borne out by relevant recent observations of a number of
our sample clusters, including NGC 2516 (cf. Sung et al. 2002), NGC
2632 (M44; also known as the Praesepe cluster: see, e.g., Bouvier et
al. 2001; Patience et al. 2002, and references therein), the Pleiades
(e.g., Martin et al. 2000, and references therein), and the Hyades
(e.g., Stefanik \& Latham 1992; Patience et al. 1998).

That the cluster cores appear to lie below the SSP predictions seems
to suggest that the effect of binaries outweighs mass segregation and
the change in the MF in determining the position of the cluster cores
in the diagram. As shown by Kouwenhoven \& de Grijs (2008), this is to
be expected for relatively low-mass open clusters such as the objects
we are considering here. However, without a detailed investigation of
each cluster and their component stars, it is impossible to
reconstruct the degree to which each effect is important. However, we
argue that it seems clear that the position of most clusters (cores)
below the canonical SSP lines is not due to significant deviations
from virial equilibirium.

Finally, we also note that we may well have overestimated the masses
by factors of a few through the universal use of
Eq. (\ref{spitzer.eq}). For highly mass-segregated clusters
containing significant binary fractions, a range of stellar IMF
representations, and for combinations of characteristic relaxation
time-scales and cluster half-mass radii, the adoption of a single
scaling factor $\eta \approx 9.75$ introduces systematic offsets,
leading to smaller values of $\eta$ (e.g., Fleck et al. 2006;
Kouwenhoven \& de Grijs 2008), and thus to dynamical mass
overestimates if $\eta = 9.75$ were assumed. However, the
uncertainties are too large at the present time to reach firm
conclusions regarding the dependence of our results on $\eta$.

\subsection{Westerlund 1}
\label{wld1.sec}

\begin{figure}
\includegraphics[width=\columnwidth]{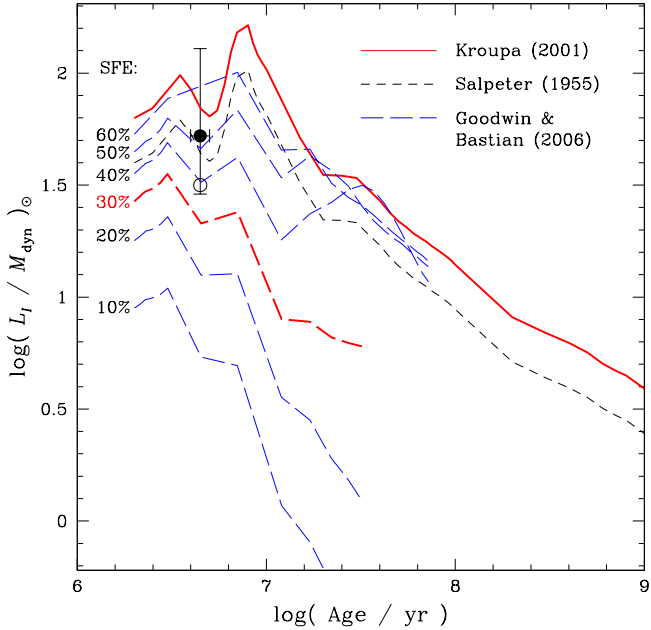}
\caption{\label{wld1.fig}Westerlund 1 in the diagnostic age versus
$M/L_V$ ratio diagram. The open circle represents the cluster's locus
if we were to exclude the nine brightest stars; this exemplifies the
uncertainties introduced by stochastic IMF sampling and by having
fortuitously caught the cluster at a time when it is dominated by a
small number of very bright stars. The line coding is as in
Fig.~\ref{ocl.fig}.}
\end{figure}

The Galactic young massive star cluster, Westerlund 1 (aged 4--5 Myr;
Crowther et al. 2006), and in particular its stellar content, has been
the subject of considerable recent attention (e.g., Clark et al. 2005,
2008; Crowther et al. 2006; Muno et al. 2006; Mengel \& Tacconi-Garman
2007a,b; and references therein). It is the nearest potential GC
progenitor, and certainly the most massive young Galactic cluster
($M_{\rm cl} \simeq 10^5 M_\odot$, with an absolute lower limit of
$M_{\rm cl,low} \simeq 1.5 \times 10^3 M_\odot$; Clark et al. 2005;
see also Mengel \& Tacconi-Garman\footnote{Although these authors
published a mass determination of $M_{\rm cl} = 6.3^{+5.3}_{-3.7}
\times 10^4 M_\odot$ for Westerlund 1, they recently redetermined its
velocity dispersion and hence its mass, at $M_{\rm cl} \sim 1.25
\times 10^5 M_\odot$ (Mengel \& Tacconi-Garman 2007b).} 2007a). In
order for the cluster to survive, it cannot have a stellar IMF that is
deficient in low-mass stars. Given that all observed star clusters
exhibit a range in stellar masses, we conclude that the Westerlund 1
IMF must therefore be close to `normal' (there is no conclusive
evidence for clusters with `bottom-heavy' IMFs, which could
potentially also lead to the cluster's position in
Fig. \ref{wld1.fig}). In de Grijs \& Parmentier (2007) we reviewed the
balance of evidence (e.g., Muno et al. 2006; Clark et al. 2008), but
the results remained inconclusive because of the difficulty of
observing the cluster's low-mass stellar population. Brandner et
al. (2008) recently completed a detailed study of the cluster's mass
function down to $\sim 1 M_\odot$, which appears to be consistent with
a normal Kroupa or Salpeter-type IMF.

In addition, the relaxation of an idealised cluster and the
contribution of the most massive stars to the escape of stars below a
typical limiting (high) mass scales approximately in a power-law
fashion (with a power $\ga 3$) with mass (e.g., H\'enon 1969, and
references therein). This is key to understanding the dynamical
importance of a particular IMF. It follows that the escape rate of
low-mass stars below a certain mass is mass-independent. Moreover, in
an equilibrium system the number of stars escaped from the cluster by
an age of 4 Myr would therefore be small and hardly affect the overall
shape of the IMF. Hence, only a small modification of the IMF below
the supernova mass limit would be expected at the present age, and the
observed IMF will therefore be close to `normal'.

Using the dynamical mass estimate from Mengel \& Tacconi-Garman
(2007a), combined with the integrated photometry of Piatti et
al. (1998), we reached a similar conclusion (de Grijs \& Parmentier
2007), despite the significant uncertainties in the observables. Since
in the $V$ band, on which the Piatti et al. (1998) photometry was
based, the confusion between the cluster members and the Galactic
field stellar population is substantial (in essence because of the
significant extinction along this sightline), we obtained imaging
observations at longer wavelengths, where this confusion is
significantly reduced. An $I$-band (peak-up) image of the cluster
(using the Ic/Iwp-ESO0845 filter), with an exposure time of 3.0~s, was
obtained with the ESO 2.2m telescope equipped with the Wide-Field
Imager (WFI) at La Silla Observatory (Chile). The image was kindly
made available to us by P. A. Crowther. Using the photometric
zero-point offsets of Clark et al. (2005) we obtained an integrated
$I$-band magnitude of $m_I = 6.15 \pm 0.05$ mag within a radius of
108~arcsec. This includes all of the bright cluster members and
excludes bright foreground sources.

The combined integrated magnitude of the three brightest red
supergiants (objects 26, 237, and 20 of Clark et al. 2005; in order of
decreasing brightness), yellow hypergiants (objects 32, 4, and 8) and
blue supergiants (objects 243, 16, and 7) is $m_I = 7.15 \pm 0.05$
mag. Therefore, these nine sources alone contribute some 40\% of the
cluster's total integrated $I$-band flux. Each of these sources is in
a rare, short-lived phase and so the luminosity of the cluster might
be expected to vary significantly on short time-scales. In addition,
we specifically discuss these nine brightest cluster members
separately, because these are the stars that make Westerlund 1 one of
the most unusual young star clusters known (e.g., Clark et al. 2005,
2008; Crowther et al. 2006). These nine sources stand out from the
overall stellar luminosity function, which appears to otherwise have
been drawn from a `normal' IMF. Thus, this serves as a clear caution
that stochasticity in the cluster's IMF (e.g., Brocato et al. 2000),
as well as stochasticity in the numbers of stars in unusually luminous
post-main-sequence evolutionary stages (e.g., Cervi\~no \&
Valls-Gabaud 2003; Cervi\~no \& Luridiana 2006) may contribute
significantly to variations in a cluster's observed $M/L$ ratio. On a
related note, we caution that the luminosities of {\em all} of the
clusters we analyse in this paper are subject to stochastic effects,
regardless of their age (Cervi\~no \& Valls-Gabaud 2003; Cervi\~no \&
Luridiana 2006).

Using the most up-to-date distance to Westerlund 1, $D = 3.9 \pm 0.7$
kpc (Kothes \& Dougherty 2007), a $V$-band extinction of $A_V = 11.6$
mag (Clark et al. 2005), and the Galactic reddening law of Rieke \&
Lebofsky (1985; resulting in $A_I = 5.6$ mag), we obtain the locus in
(age versus $L/M$ ratio) space as shown in Fig.~\ref{wld1.fig}.
Despite the large error bars and very young age, it appears at first
sight that Westerlund~1 is not significantly out of virial
equilibirum. Its location in Fig. \ref{wld1.fig} is consistent with
the cluster having formed with a high eSFE, and with a Kroupa or
Salpeter-like stellar IMF. Given the different filters used between
Piatti et al. (1998) and this paper, and in view of the updated
cluster mass estimate, this result confirms our earlier assertion in
de Grijs \& Parmentier (2007) based on the cluster's location in the
$M/L_V$ versus age diagram (which in turn supported the conclusion of
Mengel \& Tacconi-Garman 2007a that the cluster appears to be close to
virial equilibrium). However, we note that it is not entirely clear if
Westerlund~1 would be expected to be in virial equilibrium. Bastian et
al. (2008) show that observations of massive young clusters suggest
that they may expand from initial half-mass radii of $\sim 0.1$~pc to
$> 1$~pc in the first few~Myr of their lives (see, in particular,
their fig.~4). In such a situation we might expect Westerlund~1 to lie
well below the canonical IMF lines.

If we were to exlude the nine brightest stars making up some 40\% of
the cluster's integrated $I$-band flux, its locus would shift to that
of the open circle (assuming that the cluster's mass remains
unchanged). We will now briefly explore whether this effect would be
significantly different in the $V$ band, as discussed in de Grijs \&
Parmentier (2007). Although we do not have unsaturated $V$-band images
of Westerlund 1, we can use the results of Piatti et al. (1998; their
fig. 8) to check the above statements to first order. We base our
analysis on the simplistic assumption that the innermost nine stars
are the nine brightest stars.

Based on this figure, the nine innermost stars contribute a combined
$M_V \sim -9.8$ mag; the full integrated cluster magnitude is
$M_{V,{\rm tot}} \sim -11.2$ mag. We therefore conclude that these
nine innermost stars contribute $\sim$30\% of the cluster's total
luminosity. Some, but not all, of these stars are clearly the very
bright stars we used in our $I$-band analysis, so that this estimate
provides a lower limit to the contribution of the nine brightest
stars. Given that we found that in the $I$ band the nine brightest
stars contribute $\sim$40\% of the total flux, the $V$ and $I$-band
contributions of the nine brightest stars are similar, particularly in
view of the uncertainties. In fact, this could have been expected --
to first order -- because the spectral energy distributions of each of
the three subgroups (blue and red supergiants and yellow hypergiants)
peak at different wavelengths. Therefore, stochasticity remains a
serious issue across these wavelengths, simply because these stars are
intrinsically so bright.

This shows the potential effects of (i) stochastic sampling of a
cluster's IMF (predominantly affecting the highest-mass end in any
cluster) and (ii) having caught the cluster at a time when it is
dominated by a few very luminous yet short-lived stars. If indeed we
are fortuitous in having observed a stochastically exceptional
situation regarding the numbers of very massive (bright) stars in the
cluster, it would indicate that Westerlund~1 may have formed with an
eSFE around the $\sim 30$--40\% required for clusters to survive the
gas expulsion phase (e.g., Lada, Margulis \& Dearborn 1984; Goodwin
1997a,b; Adams 2000; Geyer \& Burkert 2001; Kroupa \& Boily 2002;
Boily \& Kroupa 2003a,b; Fellhauer \& Kroupa 2005; Bastian \& Goodwin
2006; Parmentier \& Gilmore 2007, their fig. 1), although we note that
the error bars are large and will remain unchanged by the removal of
these nine brightest stars. As an aside, we note that differential
extinction towards the individual brightest cluster stars is not an
issue; extinction variations along individual sight lines are minimal,
with deviations from a mean $K_{\rm s}$-band extinction of
$\Delta(A_{K_{\rm s}}) = 0.14$ mag (e.g., Crowther et al. 2006); this
small spread is due to the extinction estimates being based on assumed
intrinisic stellar colours, which in reality vary slightly.

\subsection{The Orion Nebula Cluster}
\label{onc.sec}

The dynamical state of the core of the Orion Nebula Cluster (NGC 1976,
M42; cluster 1 in Fig. \ref{ocl.fig}) has been the subject of
significant observational and theoretical investigations (e.g.,
Hillenbrand \& Hartmann 1998; Kroupa, Petr \& McCaughrean 1999; Kroupa
2000; Kroupa et al. 2001; O'Dell 2001; Scally, Clarke \& McCaughrean
2005; and references therein). It is the youngest cluster in our
sample and is located ($\sim$3$\sigma$) below the `normal' SSP
evolution in Fig. \ref{ocl.fig}, even in view of the
uncertainties. This super-virial state is corroborated by current
estimates of its virial ratio, which suggest that the cluster (core)
is already unbound, but has only recently become so (e.g., Kroupa et
al. 2001; Scally et al. 2005). In fact, Hillenbrand \& Hartmann (1998)
showed that in order for the ONC to be in virial equilibrium, based on
the cluster's observed velocity dispersion, the total mass within
about 2 pc of the central `Trapezium' configuration of massive stars
must be of order twice that of the known stellar population in the
region (and comparable to the estimated mass in molecular gas
projected onto the area). Given the youth of the cluster, and its
partially embedded nature, Hillenbrand \& Hartmann (1998) argued that
if $\ga 20$\% of the remaining molecular gas is converted into stars,
this might result in a gravitationally bound cluster. Follow-up
$N$-body simulations led Scally et al. (2005) to conclude that the
size and age of the ONC imply that either the cluster is marginally
bound (or has become unbound only very recently), or else that it has
expanded quasi-statically. Kroupa et al. (2001), on the other hand,
performed binary-rich {\it N}-body models of the ONC adopting two of
the allowed initial configurations from Kroupa (2000) and showed that
it is currently expanding and was probably formed with an eSFE near
33\%. In view of the uncertainties, this is roughly consistent with
its locus in the diagnostic diagram of Fig. \ref{ocl.fig}.

\subsection{Dissolving clusters}
\label{dissolve.sec}

\subsubsection{Coma Berenices}

Odenkirchen (1998) found that the open cluster in Coma Berenices
(cluster 10 in Fig. \ref{ocl.fig}) has an elliptical core-halo
morphology, combined with a group of extratidal stars (either escaping
stars or genuine field stars; see also K\"upper et al. 2008 for a
general discussion on the distribution of escaped stars), which are
located at projected distances of $\ge 10$ pc from the cluster
centre. They provide some tentative evidence for the presence of an
additional population of even lower-mass extratidal stars, and argue
that the existence of this significant population of stars beyond the
cluster's tidal radius is evidence of the cluster dissolution process
caught in the act. At the same time, they conclude that -- given the
present mass and configuration of the cluster stars -- the observed
(core) velocity dispersion is fully consistent with the expectations
from the SSP models.

Here we reach, in essence, the same conclusion. Based on the
observational data at hand, the core of the star cluster in Coma
Berenices is located very close to the expected photometric
evolutionary sequences in Fig. \ref{ocl.fig}, within reasonably small
uncertainties. Given that there is evidence that this cluster is in
the advanced stages of dissolution, this result should be considered
as a strong caution. It appears that for a cluster to survive for a
significant length of time, it is a necessary, but {\it not a
sufficient condition} for it to be located close to the evolutionary
sequences in our diagnostic diagram. We caution, however, that since
our velocity dispersion measurements were weighted towards the central
region of the cluster, it is possible that the cluster's locus in
Fig. \ref{ocl.fig} mainly reflects its remaining bound component.

\subsubsection{The Hyades}

The Hyades (cluster 11 in Fig. \ref{ocl.fig}) is a dynamically very
evolved, marginally bound cluster significantly depleted in low-mass
stars (e.g., Kroupa 1995; Perryman et al. 1998; see also Portegies
Zwart et al. 2001), with a stellar velocity dispersion on the order of
0.3--0.4 km s$^{-1}$ (Makarov, Odenkirchen \& Urban 2000; see also
Madsen 2003). Detailed $N$-body simulations (e.g., Terlevich 1987;
Madsen 2003; Chumak, Rastorguev \& Aarseth 2005) indicate that a halo
of gravitationally unbound stars can still be linked with the cluster,
and that these stars are moving along with it on similar orbits
(cf. K\"upper et al. 2008), for several $\times 10^8$ yr (see Perryman
et al. 1998 for a detailed discussion). Hence, at its current age of
$\log t ({\rm yr}) = 8.85^{+0.08}_{-0.09}$ (Paunzen \& Netopil 2006;
and references therein) it is not surprising that the Hyades moving
group is still detectable as a cluster-type object.

In Fig. \ref{ocl.fig}, the Hyades occupies a locus very close to the
evolutionary sequences (and with small error bars), yet the group is
likely (i) unbound overall and (ii) in the final stages of dissolution
(Odenkirchen et al. 1998). Although the same caution applies to the
Hyades moving group as to the cluster in Coma Berenices, we conclude
again that for a cluster to survive for a significant length of time,
it is a necessary, but {\it not a sufficient condition} for it to be
located close to the evolutionary sequences in our diagnostic diagram.

\section{Discussion and conclusions}
\label{summary.sec}

In this paper, we have explored the usefulness of the diagnostic age
versus $M/L$ ratio diagram in the context of Galactic open clusters.
This diagram is often used in the field of extragalactic young to
intermediate-age massive star clusters to constrain the shape of their
stellar IMF, as well as their stability and the likelihood of their
longevity.

Using a sample of Galactic open clusters for which reasonably accurate
internal (core) velocity dispersions are available in the literature,
we constructed a homogenised set of observational data drawn from a
wide variety of publications, also including their most likely
uncertainty ranges. This allowed us to derive dynamical mass estimates
for our sample of open clusters, as well as their respective $M/L_V$
ratios and -- crucially -- the associated (realistic) uncertainties.

It seems clear that the effect of binaries, mass segregation, and the
dynamical alteration of mass functions by two-body relaxation are
important constraints that cannot be ignored.

Using the massive young Galactic cluster Westerlund~1 as a key
example, we caution that stochasticity in the IMF introduces
significant additional uncertainties. Therefore, the stability and
long-term survival chances of Westerlund~1 remain inconclusive.

Most importantly, however, we conclude that for an open cluster to
survive for any significant length of time (in the absence of
substantial external perturbations), it is a necessary but not a
sufficient condition to be located close to the predicted photometric
evolutionary sequences for `normal' SSPs. This is highlighted using a
number of our sample clusters (and the parameters related to the
cluster cores) which are known to be in a late stage of dissolution,
and lie very close indeed to either of the evolutionary sequences
defined by the Salpeter (1955) or Kroupa (2001) IMFs.  However, we
also note that a fair fraction of our sample clusters show the
signatures of dynamical relaxation and stability. Among our current
sample, these include NGC 2168 (M35; Kalirai et al. 2003), NGC 2682
(M67; Hurley et al. 2005), NGC 6705 (M11; Mathieu 1984; McNamara \&
Sekiguchi 1986; Sung et al. 1999) and the Pleiades (M45; McNamara \&
Sekiguchi 1986; Pinfield, Jameson \& Hodgkin 1998; Raboud \&
Mermilliod 1998). Despite their relatively small masses ($M_{\rm cl}
\la 2 \times 10^3 M_\odot$) and ages in excess of a few $\times 10^8$
yr, this is not unexpected.

Using the vertical oscillation period, $\pi$, around the Galactic
plane of NGC 2323 ($\pi \approx 50$ Myr; Clari\'a et al. 1998) as an
example, this cluster has only been through a few of these periods,
given its age of $\log t ({\rm yr}) = 8.11^{+0.05}_{-0.25}$ (Kalirai
et al. 2003). However, at the Galactocentric distance of the Sun, a
Pleiades-like open cluster crosses the Galactic disc approximately
10--20 times before it dissolves (de la Fuente Marcos 1998a,b). The
models of Kroupa et al. (2001), which match the ONC at an age of 1 Myr
very well, as well as the Pleiades at 100 Myr, suggest that these
objects would end up below the evolutionary sequences in
Fig. \ref{ocl.fig}, despite having started from the Kroupa (2001) IMF
at birth. The deviation may have been caused by the heating of the
clusters by the Galactic tidal field. In other words, it seems that
the velocity dispersion is always somewhat higher after Galactic-plane
passage, because the stars suffer from an additional acceleration.

Similarly, the age of NGC 2516 (cluster 3 in Fig. \ref{ocl.fig}),
$\log t ({\rm yr}) = 8.2 \pm 0.1$ (Sung et al. 2002), is well in
excess of its period of vertical oscillations through the Galactic
plane, $\pi \simeq (7-8) \times 10^7$ yr (Dachs \& Kabus 1989). NGC
2516 is presently located some 120 pc below the Galactic plane, near
the dense molecular clouds of the Vela Sheet. These Galactic plane
passages may have contributed to rendering the cluster unstable.
Alternatively, encounters with giant molecular clouds (e.g., Gieles et
al. 2006), particularly around the time of Galactic plane passages,
may have contributed to the cluster's present dynamical state.

In addition, Bonatto \& Bica (2003) show that tidal losses of stars
from NGC 2682 to the Galactic field have been effective (see also
McNamara \& Sekiguchi 1986 for, e.g., NGC~2168 and NGC 2632). This
interpretation is supported by the $N$-body simulations of Hurley et
al. (2005). Additional (circumstantial) evidence for tidal effects
acting on NGC 2682 is present in the form of significantly elliptical
cluster isophotes (Fan et al. 1996), which might be a tidal extension
caused by the Galactic field (Bergond, Leon \& Guibert
2001).\footnote{NGC 3532 is also strongly flattened (Gieseking 1981),
roughly orthogonal to the Galactic plane. Both theory and $N$-body
simulations suggest that the effects of the Galactic tidal field give
rise to a flattening of cluster outskirts in the direction towards the
Galactic Centre and perpendicular to the Galactic plane (see Mathieu
1985).} In addition, Chupina \& Vereshchagin (1998) detected several
density enhancements in the low-density extended outskirts of the
cluster. Such clumps are expected as a consequence of disc shocking
(e.g., Bergond et al. 2001). Alternatively, the cluster may have
undergone a number of encounters with giant molecular clouds, possibly
leading to a similar morphology. Similarly, Adams et al. (2001)
suggest that the flattening of the stellar mass function of the
Pleiades below $m_\ast \sim 0.2 M_\odot$ with respect to the
field-star population may have been caused by evaporation of the
lowest-mass stars into the Galactic field (see also van Leeuwen 1983),
although evidence for this to be the case remains inconclusive (see,
e.g., the simulations of de la Fuente Marcos 2000; Moraux, Kroupa \&
Bouvier 2004).

In a follow-up paper (Kouwenhoven et al., in prep.) we will
quantitatively explore the loci in the diagnostic diagram of the
Galactic open clusters, using $N$-body simulations.

\section*{Acknowledgments} We acknowledge research support and
hospitality at the International Space Science Institute in Bern
(Switzerland), as part of an International Team programme. We thank
Paul Crowther for discussions and the use and basic analysis of his
unpublished imaging of Westerlund 1, which was obtained as part of ESO
proposal 69.D-0327(B). We are particularly grateful to Mark Gieles for
very helpful comments. We also thank Vladimir Danilov for providing us
with some hard-to-find data for a few of our sample clusters and
acknowledge the constructive criticism in the referee report of Sverre
Aarseth. RdG and MK acknowledge financial support from STFC grant
PP/D002036/1; RdG also acknowledges partial support from the Royal
Society in the context of a `Frontiers of Science' programme. This
research has made use of the SIMBAD database, operated at CDS,
Strasbourg (France), and of NASA's Astrophysics Data System Abstract
Service.

\end{document}